\documentclass{cimento}

\title{Gravitating macroscopic media in general relativity\\ 
and macroscopic gravity}
\author{Giovanni~Montani\from{icra}\thanks{E-mail address:
montani@vxrmg9.icra.it},
        Remo~Ruffini\from{icra}\thanks{E-mail address: ruffini@icra.it}
        \atque Roustam~Zalaletdinov\from{icra}\from{iyaf}\thanks{E-mail 
address: zala@icra.it}}
\instlist{\inst{icra} ICRA, Departamento di Fisica, Universit\'a di Roma 
``La Sapienza" P.le Aldo Moro 5, Roma 00185, Italia
          \inst{iyaf} Department of Theoretical Physics, Institute of 
Nuclear Physics, Uzbek Academy of Sciences, Tashkent 702132, Uzbekistan, CIS}
\PACSes{\PACSit{04.20. q}{Classical general relativity}
        \PACSit{04.90.+e}{Other topics in general relativity and gravitation}
        \PACSit{83.20.Bg}{Macroscopic (phenomenological) theories}}
\begin{document}

\maketitle

\begin{abstract}
The problem of construction of a continuous (macroscopic) matter model for 
a given point-like (microscopic) matter distribution in general relativity 
is formulated. The existing approaches are briefly reviewed and a physical
analogy with the similar problem in classical macroscopic electrodynamics is 
pointed out. The procedure by Szekeres in the linearized general relativity 
on Minkowski background to construct a tensor of gravitational quadruple 
polarization by applying Kaufman's method of molecular moments for derivation 
of the polarization tensor in macroscopic electrodynamics and to derive an 
averaged field operator by utilizing an analogy between the linearized 
Bianchi identities and Maxwell equations, is analyzed. It is
shown that the procedure has some inconsistencies, in particular, 
(1) it has only provided the terms linear in perturbations for the averaged 
field operator which do not contribute into the dynamics of the averaged
field, and (2) the analogy between electromagnetism and gravitation does 
break upon averaging. A macroscopic gravity approach in the perturbation 
theory up to the second order on a particular background space-time taken 
to be a smooth weak gravitational field is applied to write down a system 
of macroscopic field equations: Isaacson's equations with a source
incorporating the quadruple gravitational polarization tensor, 
Isaacson's energy-momentum tensor of gravitational waves and energy-momentum 
tensor of gravitational molecules and corresponding equations of motion. A 
suitable set of material relations which relate all the tensors is proposed. 
\end{abstract}

\section{Introduction}
\label{intro}
In using the Einstein equations for a matter distribution
in the form of a set of point-like mass constituents, there is a problem of 
adequate application, or validity, of the Einstein equations when 
such a matter distribution is substituted by a continuous matter
distribution while the field operator in 
the left-hand side of the equations is kept unchanged. 
This problem as it stands in cosmology\footnote
{For discussion on the other physical settings on general relativity 
facing the same problem see~\cite{Tava-Zala:1998}.} 
is called {\em the averaging 
problem}~\cite{Shir-Fish:1962,Scia:1971,Elli:1984,Zala:1992,Zoto-Stoe:1992}. 
Indeed, let us consider the Einstein
equations in the mixed form\footnote{The mixed form is preferable here for 
the reason that it contains only products of metric by curvature. On contrary, 
the covariant or contravariant forms of the Einstein equations have triple 
products of metric by metric by curvature.}
\begin{equation}
\label{EE} 
g^{\alpha \epsilon }r_{\epsilon \beta }-\frac 12\delta _\beta
^\alpha g^{\mu \nu }r_{\mu \nu }=-\kappa t_\beta ^{\alpha {\rm (discrete)}} 
\end{equation}
with
\begin{equation}
\label{dmd} 
t_\beta ^{\alpha {\rm (discrete)}} (x) = 
\sum_{i}^{} {t_{(i)}}_\beta^{\alpha} [x-z_{(i)}(\tau_{(i)})] 
\end{equation}
where ${t_{(i)}}_\beta^{\alpha}$ is a energy-momentum tensor for a point-like 
mass moving along its world line $z^\mu = z^\mu_{(i)}(\tau_{(i)})$ parameterized 
by $\tau_{(i)}$ and $i$ counts for the matter particles in the distribution 
(\ref{dmd}). Changing the discrete matter distribution to a continuous 
(hydrodynamical) one in the right-hand side of (\ref{EE}), which is the standard 
approach in 
cosmology~\cite{Shir-Fish:1962,Scia:1971,Elli:1984,Zala:1992,Zoto-Stoe:1992} made 
phenomenologically on the basis of assumption about the uniformity and isotropicity 
of distribution of galaxies, or cluster of galaxies, throughout the whole Universe,
means an implicit averaging denoted here by $\langle \cdot \rangle $
\begin{equation}
\label{aver-dmd} 
t_\beta ^{\alpha {\rm (discrete)}} (x) \rightarrow 
T_\beta ^{\alpha {\rm (hydro)}} (x) =
\left \langle \sum_{i}^{} {t_{(i)}}_\beta^{\alpha} [x-z_{(i)}(\tau_{(i)})] 
\right \rangle \ .
\end{equation}
Given a covariant averaging
procedure $\langle \cdot \rangle $ for tensors on space-time, 
the averaging out of (\ref{EE}) with taking into account (\ref{aver-dmd}) brings 
\begin{equation}
\label{averEE:1} \langle g^{\alpha \epsilon }r_{\epsilon \beta }\rangle
-\frac 12\delta _\beta ^\alpha \langle g^{\mu \nu }r_{\mu \nu }\rangle
=-\kappa T_\beta ^{\alpha {\rm (hydro)}} \ .
\end{equation}
An important point regarding the averaged equations (\ref{averEE:1}) is that 
in this form they are just 
algebraic relations between components of the smoothed hydrodynamical 
energy-momentum 
tensor and the average products of the metric tensor by the Ricci tensor 
$\langle g^{\alpha \epsilon }r_{\epsilon \beta }\rangle$ and cannot therefore 
be taken as field equations. By splitting the products out 
as $\langle g^{\alpha \epsilon }r_{\epsilon \beta }\rangle =
\langle g^{\alpha \epsilon }\rangle \langle r_{\epsilon \beta
}\rangle + C^{\alpha}_{\beta }$ where $C^{\alpha}_{\beta }$ is a
correlation tensor, the averaged equations (\ref{averEE:1}) become
\begin{equation}
\label{averEE:2} \langle g^{\alpha \epsilon }\rangle \langle r_{\epsilon
\beta }\rangle -\frac 12\delta _\beta ^\alpha \langle g^{\mu \nu }\rangle
\langle r_{\mu \nu }\rangle = 
-\kappa T_\beta ^{\alpha {\rm (hydro)}} 
- C^{\alpha}_{\beta } + \frac 12\delta _\beta ^\alpha C^{\epsilon}_{\epsilon }. 
\end{equation}
Here $\langle g^{\alpha \beta }\rangle $ and $\langle r_{\alpha \beta
}\rangle $ denote the averaged inverse metric and the Ricci tensors which 
are supposed to describe the gravitational field due to the matter
distribution $T_\beta ^{\alpha {\rm (hydro)}}$. A simple important
observation~\cite{Zala:1997,Zala:1998} now is that the averaged Einstein 
equations (\ref{averEE:2}) are still
not ``real" field equations - just a definition of the correlation tensor
$C^{\alpha}_{\beta}$ as a difference between (\ref{averEE:1}) and 
(\ref{averEE:2}). The origin of this fundamental fact is that the average 
of the non-linear operator 
of (\ref{EE}) on the metric tensor $g_{\rho \sigma}$ is not equal 
in general\footnote{It should noted that the inequality (\ref{aver:oper})
has been observed in all possible averaging settings, for example, for 
a volume space-time averaging in~\cite{Shir-Fish:1962,Zala:1992}, 
in the framework of a kinetic approach in~\cite{Yodz:1971} and for a 
statistical ensemble averaging in~\cite{Igna:1978}. Relations between 
different averaging 
procedures are discussed in~\cite{Zala:1997,Zala:1998}.}
to an operator of {\em the same form} on the average metric 
$\langle g_{\rho \sigma} \rangle$:
\begin{equation}
\label{aver:oper} 
\left \langle \left( g^{\alpha \epsilon } r_{\epsilon \beta }
- \frac 12\delta _\beta ^\alpha g^{\mu \nu }
r_{\mu \nu }  \right) [g_{\rho \sigma}] \right \rangle \neq
\left( \langle  g^{\alpha \epsilon } \rangle 
\langle r_{\epsilon \beta } \rangle 
- \frac 12\delta _\beta ^\alpha \langle g^{\mu \nu } \rangle 
\langle r_{\mu \nu } \rangle \right) [\langle g_{\rho \sigma} \rangle] \ .
\end{equation}
In order to return them the status of the field equations one must define 
the object $C^{\alpha}_{\beta}$ and find its properties using information 
outside the Einstein equations. 

To resolve the averaging problem, and to consider it in a broader context 
as the problem of macroscopic description of gravitation, the approach of 
macroscopic gravity has 
been proposed~\cite{Zala:1992,Zala:1997,Zala:1998,Zala:1993,Zala:1996,Kras:1997,
Mars-Zala:1997,Zala:1996b} (see~\cite{Zala:1997,Zala:1998,Kras:1997} 
for discussion of the problem and references therein,~\cite{Tava-Zala:1998} 
for discussion of the physical
status of general relativity as either a microscopic or macroscopic 
theory of gravity).
A covariant space-time volume averaging procedure for tensor 
fields~\cite{Zala:1992,Zala:1993,Mars-Zala:1997},
has been defined and proved to exist on arbitrary Riemannian space-times with
well-defined properties of the averages. Upon utilizing the averaging scheme, 
the macroscopic gravity approach has shown that ({\em i}) averaging out Cartan's 
structure equations brings about the structure equations for the averaged 
(macroscopic) non-Riemannian geometry and the definition and the properties 
of the correlation tensor $C^{\alpha}_{\beta}$, ({\em ii}) the averaged
Einstein's equations (\ref{averEE:2}) become then the macroscopic field 
equations and they must be supplemented by a set of differential equations 
for the correlation
tensor, ({\em iii}) it is always possible to extract the field operator 
of the form (\ref{EE}) for the Riemannian macroscopic metric tensor 
$G_{\mu \nu}$ and its Ricci tensor $M_{\mu \nu}$ with all other 
non-Riemannian correlation terms going to the right-hand side of 
(\ref{averEE:2}) to give geometric correction to the averaged 
energy-momentum tensor $T_\beta ^{\alpha {\rm (hydro)}}$. It is been also 
shown~\cite{Zala:1997,Zala:1998,Zala:1996b} that 
only in case of neglecting all correlations of the gravitational field 
the averaged equations (\ref{averEE:2}) becomes the macroscopic Einstein 
equations with a continuous matter distribution
\begin{equation}
\label{macroEE} 
G^{\alpha \epsilon }M_{\epsilon \beta }-\frac 12\delta _\beta^\alpha 
G^{\mu \nu }M_{\mu \nu }=-\kappa T_\beta^{\alpha {\rm (hydro)}} \ ,
\end{equation}
which reveals the physical status of using the standard procedure in 
cosmology of claiming (\ref{averEE:2}) to be the Einstein equations 
(\ref{macroEE}) after substitution of the matter model (\ref{aver-dmd}). 
The physical meaning,
dynamical role and magnitude of the gravitational correlations must be
elucidated in various physical settings. There is some evidence that they cannot
be negligible for cosmological evolution (see, for example~\cite{Bild-Futa:1991} 
for an estimation of the age of Universe in a second order perturbation approach).

\section{Macroscopic media in general relativity}
\label{media}
Derivation of the macroscopic (averaged) Maxwell field operator in macroscopic 
electrodynamics is easily accomplished due its linear
field structure and 
the main problem consists in the construction of models of macroscopic  
electromagnetic media (for example, diamagnetics, magnetics, waveguides, 
etc.)~\cite{Lore:1916,Pano-Phil:1962,deGr-Sutt:1972,Jack:1975}, 
which relates to the structure of the averaged
current. In general relativity the problem of construction of macroscopic
gravitating medium models is hardly elaborated due to the following reasons: 
(a) existing mathematical and physical difficulties in establishing
the form of the averaged (macroscopic) operator in (\ref{averEE:2}) for 
the field equations of macroscopic gravity recedes the interest in 
development of macroscopic gravitating media; (b) posing on its own more or 
less realistic problem with a discrete matter creates mathematical and physical 
problems due to nonlinearity and non-trivial geometry of gravitation, to mention, 
for example, the $N$-body problem, the problem of statistical description of gravity, 
etc. and (c) being relied on physically motivated phenomenological arguments 
(uniformity, isotropy, staticity, etc.) most applications of general relativity 
deal with 
{\em effective} continuous media if even a starting physical model is discrete 
in its nature like in cosmology (see Section \ref{intro}) or in description of 
extended bodies in general relativity (see~\cite{Tava-Zala:1998} for discussion of 
the physical status of general relativity).

The kinetic approach in physics is known to provide a general scheme for 
introduction of
characteristics of continuous media with a {\em known} distribution function of
a discrete configuration. But the advantages of such generality are often 
greatly weakened in particular applications by difficulties of solving the 
Boltzmann 
equation to find a distribution function of interest. This applies to
a great extent to general relativity where despite the formulation of the
general relativistic Boltzmann equation~\cite{Cher:1962,Isra:1972}
the kinetic approach still remains useful for general definitions and
considerations rather than being a working tool (see, for 
example,~\cite{Yodz:1971})
to derive a specific model of a macroscopic medium. 

In case of the macroscopic electrodynamics together with the volume space-time
averaging on Minkowski space-time the formalism of statistical 
distribution functions has been utilized (see~\cite{deGr-Sutt:1972} 
and references therein)
and it is of importance for the mathematically well-posed derivation of 
the macroscopic
theory and the general structure of averaged current starting 
from microscopic electrodynamics of point-like moving charges.
Further application of the macroscopic theory requires usually mainly 
phenomenological considerations to establish material relations between 
macroscopic average fields and induction field necessary to make an 
overdetermined system of macroscopic equations determined. A correct derivation 
of material relations
is known to require~\cite{deGr:1969} averaging the microscopic equations with a 
given microscopic matter model during accomplishing an averaging of the microscopic 
field equations. Though it is the only self-consistent way, the elaboration of such 
kind of approach still remains a challenge even for simple physical settings.

On the other hand, volume (space, time, space-time) averaging procedures 
maintain their importance, direct physical meaning in application
to macroscopic settings and their extreme clearness and descriptiveness. 
A volume averaging is also known to be unavoidable in all macroscopic
settings (including statistical 
approaches)~\cite{Jack:1975,deGr:1969,Russ:1970,Robi:1973} 
and space-time averages of physical fields are 
known to have the physical meaning as directly measurable 
quantities~\cite{Bohr-Rose:1933,DeWi:1962} (for discussion 
see~\cite{Zala:1997}, 
\cite{Zala:1998} and references therein). That greatly motivates 
and supports interest in
applying approaches with various averaging schemes in physics despite
corresponding (mostly mathematical, not physical) difficulties in the 
rigorous formulation of 
averaging procedures. 

The paper aims to approach the problem of construction of gravitating macroscopic
media in general relativity by using an appropriate space-time averaging scheme.  

\section{Szekeres' gravitational polarization tensor}

The approach of Szekeres~\cite{Szek:1971} to construct a tensor 
of gravitational quadruple
polarization is based on the following assumptions:
(a) the linearized theory of gravity on Minkowski background;
(b) the linearized field equations are taken as the linearized Bianchi
identity to employ an analogy between gravitation and electromagnetism;
(c) the covariant method of molecular moments of Kaufman is applied to
construct a tensor of quadruple gravitational polarization.

The equations under consideration are the contracted Bianchi identities 
\begin{equation}
\label{bianchi}
C_{\mu \nu \rho \sigma}{}^{; \sigma} = \kappa J_{\mu \nu \rho}
\end{equation}
where $C_{\mu \nu \rho \sigma}$ is the Weyl tensor interpreted as 
free gravitational 
field~\cite{Pira:1962,Szek:1966}, $J_{\mu \nu \rho}$ is a kind of 
``matter current" for the energy-momentum tensor $t_{\mu \nu}$  
\begin{equation}
\label{matter-current}
J_{\mu \nu \rho} = J_{[ \mu \nu ] \rho} = - (t_{\rho [ \mu ; \nu ]} - 
\frac{1}{3} g_{\rho [ \mu} t_{, \nu ]}) ,
\end{equation}
\begin{equation}
\label{matter-current-conserv}
J_{\mu \nu \rho}{}^{; \rho} = 0 .
\end{equation}
equations (\ref{bianchi}) are analogous the Maxwell equations 
\begin{equation}
\label{maxwell}
f_{\mu \nu }{}^{; \nu} = \frac{4 \pi}{c} j_{\mu} 
\end{equation}
with (\ref{matter-current-conserv}) being comparable with the 
conservation of the 
electromagnetic current $j_{\mu}$ 
\begin{equation}
\label{em-current-conserv}
j_{\mu}{}^{; \mu} = 0 .
\end{equation}

Let us consider a number of particles labeled by $i$ and having masses 
$m_i$ and which are moving in there own effective gravitational field 
along world lines
$z^{\mu}_i (\tau_i)$. A physical parameter which characterizes such a 
distribution is 
a typical characteristic distance $l$ between neighbouring particles. 
Then the corresponding microscopic energy-momentum tensor has the form
\begin{equation}
\label{micro}
t^{{\rm (micro)} \mu \nu} (x) = c^{-1} \sum_{i} \int m_i 
\frac{d z^{\mu}_i }{d\tau_i } 
\frac{d z^{\nu}_i }{d\tau_i } \delta^4 [x - z^{\mu}_i (\tau_i)] d\tau_i .
\end{equation}
Assume now that due to gravitation the particles form into groups, 
a kind of gravitational molecules, which will be labeled by index $a$. 
From the physical point of view that means the presence of another 
parameter which is a characteristic size (diameter) $L$ of such a 
molecule  with $L \gg l$. It is a longwave {\em macroscopic} parameter 
and its presence in a 
microscopic system will be defining the dynamics of the system on the 
distances of order of 
$L$. The microscopic energy-momentum tensor (\ref{micro}) becomes now 
\begin{equation}
\label{molec}
t^{{\rm (molec)} \mu \nu} (x) = c^{-1} \sum_{a} \sum_{i}^{in a} \int{}{} m_i 
\frac { d\tau_a }{ d\tau_i} 
\frac{d z^{\mu}_i }{ d\tau_a} 
\frac{d z^{\nu}_i }{d\tau_a} \delta^4 [x - y^{\mu}_a (\tau_a) - s_i(\tau_a)] d\tau_a
\end{equation}
where $y^{\mu}_a (\tau_a)$ is a world line of the $a$-th molecule center 
of mass~\cite{Szek:1971} and   
$s^{\mu}_i(\tau_a) = z^{\mu}_i (\tau_i) - y^{\mu}_a (\tau_a)$ is a vector 
connecting 
the $i$-th particle with the center of mass of the molecule including this 
particle.
Let us apply now the method of molecular moments of Kaufman~\cite{Kauf:1962} 
to represent (\ref{molec}) as a series expansion in powers of $s^{\mu}_i$ under 
assumption that the effective gravitational field which is created by moving 
gravitational molecules is a weak field and the perturbations of the gravitational 
field due to relative 
oscillations of gravitating particles in molecules are small compared with the 
mean effective field. After averaging out over a typical size of the gravitational 
molecule (for an averaging procedure see~\cite{Zala:1992,Zala:1993}) 
one gets\footnote{
No explicit averaging procedure had been used in~\cite{Szek:1971} and averaged 
relations and equations were being written rather on the basis of heuristic 
considerations than 
a rigorous analysis.} in accordance with the Szekeres procedure 
\begin{equation}
\label{av-molec}
\langle t^{{\rm (molec)} \mu \nu} \rangle = T^{{\rm (free)}\mu \nu} + 
D^{\mu \nu \rho}{}_{, \rho} + Q^{\mu \nu \rho \sigma}{}_{, \rho \sigma}
\end{equation}
where $T^{{\rm (free)} \mu \nu}$ is the energy-momentum tensor of molecules, 
which has the form similar to (\ref{micro}) with substitution $i$ by $a$, 
$D^{\mu \nu \rho}$ is the tensor of gravitational dipole polarization that can be
incorporated into the quadrupole term, and $Q^{\mu \nu \rho \sigma}$ the tensor of 
gravitational quadrupole polarization 
\begin{equation}
\label{polar}
Q^{\mu \nu \rho \sigma} = c^{-1} \langle \sum_{a} \int q^{\mu \nu \rho \sigma}_a 
\delta^4 (x - y_a) d\tau_a \rangle ,
\end{equation}
which has the symmetries of the Riemann tensor.
The expression for the covariant gravitational quadrupole moment 
$q^{\mu \nu \rho \sigma}_a$ is defined as 
\begin{equation}
\label{moment}
q^{\mu \nu \rho \sigma}_a = g^{\mu \nu}_a u^{\rho}_a u^{\sigma}_a - 
g^{\rho \nu}_a u^{\mu}_a 
u^{\sigma}_a - g^{\mu \sigma}_a u^{\rho}_a u^{\nu}_a + 
g^{\rho \sigma}_a u^{\mu}_a u^{\nu}_a +
u^{\mu}_a h^{\rho \nu \sigma}_a - u^\rho_a h^{\mu \nu \sigma}_a + 
u^\nu_a h^{\sigma \mu \rho }_a -
u^\sigma_a h^{\nu \mu \rho}_a + k^{\mu \nu \rho \sigma}_a ,
\end{equation}
where
\begin{equation}
g^{\mu \nu}_a = \sum_{i} m_i \frac{d\tau_a }{ d\tau_i} s^\mu_i s^\nu_i ,
\end{equation}
\begin{equation}
h^{\mu \nu \rho}_a = \frac{2}{3} \sum_{i} m_i \frac{d\tau_a}{d\tau_i} s^\mu_i 
\left( \frac{ds^\nu_i}{d\tau_a} s^\rho_i - 
\frac{ds^\rho_i}{d\tau_a} s^\nu_i \right) ,
\end{equation}
\begin{equation}
k^{\mu \nu \rho \sigma}_a = \frac{2}{3} \sum_{i} m_i \frac { d\tau_a }{ d\tau_i}  
\left( \frac{ds^\mu_i}{d\tau_a} s^\rho_i \frac{ds^\nu_i}{d\tau_a} s^\sigma_i - 
\frac{ds^\mu_i}{d\tau_i} s^\rho_i s^\nu_i \frac{ds^\sigma_i}{d\tau_a} \right) .
\end{equation}
Upon averaging (\ref{bianchi}) over the typical size of gravitational 
molecule the following equations were obtained:
\begin{equation}
\label{av-bianchi}
{\langle C_{\mu \nu \rho \sigma}{} \rangle} ^{, \sigma} = 
\kappa \langle J^{\rm (micro)}_{\mu \nu \rho} \rangle
\end{equation}
where
\begin{equation}
\label{av-matter-current}
\langle J^{\rm (micro)}_{\mu \nu \rho} \rangle =  
- \langle t^{\rm (micro)}_{\rho [ \mu} \rangle_{, \nu ]} + 
\frac{1}{3} \eta_{\rho [ \mu} \langle t^{\rm (micro)} \rangle_{, \nu ]} ,
\end{equation}
or \begin{equation}
\label{polarization}
P_{\mu \nu \rho \sigma} = \frac{1}{2} (- {Q_{\rho \sigma \epsilon [ \mu}}^{, \epsilon} - 
\frac{1}{3} \eta_{\rho [ \mu} 
Q^{\gamma}{}_{\sigma \gamma \epsilon}{}^{, \epsilon})_{, \nu ]} 
\end{equation}
and
\begin{equation}
\label{av-matter-current-2}
\langle J^{\rm (micro)}_{\mu \nu \rho} \rangle = J^{\rm (free)}_{\mu \nu \rho} - 
P_{\mu \nu \rho \sigma}{}^{, \sigma} . 
\end{equation}
The expression (\ref{av-matter-current-2}) is analogous to the expression for the 
averaged electromagnetic current $\langle j^{\rm (micro) \mu} \rangle$  for
a bunch of charged particles moving along their world lines in the effective 
electromagnetic
field in accordance with the microscopic equation (\ref{maxwell}) 
with $j^{\rm (micro) \mu}$
when particles are grouped into molecules~\cite{Kauf:1962}
\begin{equation}
\label{av-current}
\langle j^{\rm (micro) \mu} \rangle = j^{\rm (free) \mu} - 
c P^{\mu \nu }{}_{, \nu} 
\end{equation}
where the polarization tensor $P^{\mu \nu }$ is defined as an average 
of the quadruple 
polarization moment of the molecules $p^{\mu \nu }_a$ 
(see~\cite{Kauf:1962,Szek:1971} for details)
\begin{equation}
\label{polarization-em}
P^{\mu \nu} = \langle \sum_a \int d\tau_a p^{\mu \nu}_a \delta^4 (x - y_a) \rangle .
\end{equation}
Then equations (\ref{av-bianchi}) can be rewritten as the macroscopic 
equations\footnote{
Gravitational macroscopic equations similar to (\ref{av-bianchi}) are 
known to have been proposed first in~\cite{Bel:1961}.}
\begin{equation}
\label{av-bianchi-2}
E_{\mu \nu \rho \sigma}{}^{, \sigma} = 
\kappa J^{\rm (free)}_{\mu \nu \rho} 
\end{equation}
for the gravitational induction tensor $E_{\mu \nu \rho \sigma}$ defined as
\begin{equation}
\label{gr-induction}
E_{\mu \nu \rho \sigma} = \langle C_{\mu \nu \rho \sigma} \rangle + 
\kappa P_{\mu \nu \rho \sigma} .
\end{equation}
The macroscopic equations (\ref{av-bianchi-2}) are analogous to the 
Maxwell macroscopic equations obtainable by means of averaging the 
microscopic equations (\ref{maxwell}) with 
$j^{{\rm (micro)} \mu}$ with taking into account (\ref{av-current})
\begin{equation}
\label{av-maxwell}
H^{\mu \nu }{}_{, \nu} = \frac{4 \pi}{c} J^{\rm (free) \mu} 
\end{equation}
for the electromagnetic induction tensor $H_{\mu \nu }$ defined as 
\begin{equation}
\label{em-induction}
H^{\mu \nu } = \langle f^{\mu \nu } \rangle  + 4 \pi P_{\mu \nu } .
\end{equation}
Unfortunately, at this point the analogy between the electromagnetism and 
gravitation which holds on the level of (\ref{bianchi}), 
(\ref{matter-current-conserv}) and (\ref{maxwell}), (\ref{em-current-conserv}) 
breaks. Indeed, the formal similarity of (\ref{av-bianchi-2}), (\ref{gr-induction}) 
and (\ref{av-maxwell}), (\ref{em-induction})
does not possess the structural analogy between averaged electromagnetism and 
gravitation: (A) the gravitational induction tensor $E_{\mu \nu \rho \sigma}$ does 
not have any more the symmetries of the Weyl tensor compared with $H_{\mu \nu }$ 
keeping the symmetries of $f_{\mu \nu }$; (B) it is constructed from the second 
derivatives of the polarization tensor $Q_{\mu \nu \rho \sigma}$ compared with 
the linear algebraic structure of the 
electromagnetic induction
tensor $H_{\mu \nu }$ in terms of the polarization tensor 
$P_{\mu \nu }$ - it is thus 
impossible to proceed with the formulation of phenomenological material 
relations between 
$E_{\mu \nu \rho \sigma}$ and $\langle C_{\mu \nu \rho \sigma} \rangle$ 
as possible 
in electromagnetism (relations between $H_{\mu \nu }$ and
$\langle f_{\mu \nu }  \rangle$, or amongst the fields  
${\bf E}$, ${\bf D}$, ${\bf B}$, ${\bf H}$ and ${\bf J}$). 

Even more important issue is that analysis of the macroscopic field equation 
(\ref{av-bianchi}) 
\begin{equation}
\label{av-bianchi-orders}
{\begin{array}[t]{c}
{\langle C_{\mu \nu \rho \sigma}{} \rangle}^{, \rho}\\ 
{\scriptscriptstyle {\cal O}(e)} 
\end{array}}
= \kappa \langle J^{\rm (micro)}_{\mu \nu \rho} \rangle = 
{\begin{array}[t]{c}
\kappa J^{\rm (free)}_{\mu \nu \rho} \\
{\scriptscriptstyle {\cal O}(1)}
\end{array}}  - 
{\begin{array}[t]{c}
P_{\mu \nu \rho \sigma}{}^{, \sigma} \\
{\scriptscriptstyle {\cal O}(e^2)}
\end{array}},
\end{equation}
where $e$ is a parameter measuring the value of deviation from the flat space, 
requires 
one to put into agreement the orders of magnitude of all quantities and reveals that 
the linearized Weyl tensor should be zero under averaging 
$\langle C_{\mu \nu \rho \sigma} \rangle = 0$. The Weyl tensor must be estimated 
in the perturbation theory up to the second order as it was done for the matter 
current $J^{\rm (micro)}_{\mu \nu \rho}$ in the left-hand side of (\ref{bianchi}). 
So considering some physical features of the polarization tensor neither the macroscopic equations (\ref{av-bianchi}), nor any other field equations had in fact 
been employed~\cite{Szek:1971}. 

On the basis of the expression
\begin{equation}
\label{av-energy-momentum}
\langle t^{\rm (micro)}_{\mu \nu} \rangle = T^{\rm (free)}_{\mu \nu} +
\frac{1}{2}Q_{\mu \rho \nu \sigma}{}^{, \rho \sigma}  
\end{equation}
the following material relations have been suggested
\begin{equation}
\label{gr-material}
Q_{i 0 j 0} = \langle G_{ij} \rangle N = \epsilon_g C_{i 0 j 0}
\end{equation}
where $N$ is the average number of molecules per unit volume, $G_{ij}$ is the quadrupole 
moment of a molecule
\begin{equation}
\label{quadruple}
G_{ij} =  \int \rho (x) \delta x_i \delta x_j d^3x  ,
\end{equation}
$\rho = \rho(x)$ is the matter density in molecules, $\delta x_i$ is a vector between
neighbouring particles of a molecule. The quantity $\epsilon_g$ has been called the 
gravitational dielectric constant and in Newtonian approximation found to be 
\begin{equation}
\label{grav-dielectric-const}
\epsilon_g = \frac{1}{4} \frac{m A^2 c^2}{\omega_0^2} N
\end{equation}
where $A$ is the average linear dimension of a typical molecule, $m$ is the 
average mass of
the molecules, $\omega_0^2$ is a typical frequency of harmonically oscillating 
particles in
molecules. 

\section{Macroscopic gravity equations}

The gravitational field created by the particles is defined by Einstein's equation
\begin{equation}
\label{ein}
g^{\alpha \epsilon }r_{\epsilon \beta }-\frac 12\delta _\beta
^\alpha g^{\mu \nu }r_{\mu \nu }=-\kappa t_\beta ^{\alpha {\rm (micro)}} .
\end{equation}
The Einstein equations (\ref{ein}) for the distribution (\ref{molec}) are of the form:
\begin{equation}
\label{ein-molec}
g^{\alpha \epsilon }r_{\epsilon \beta }-\frac 12\delta _\beta
^\alpha g^{\mu \nu }r_{\mu \nu }=-\kappa t_\beta ^{\alpha {\rm (molec)}} .
\end{equation}

Averaging the left-hand side of the Einstein equations (\ref{ein-molec}) 
following the 
Isaacson's high-frequency approximation approach~\cite{Isaa:1968a,Isaa:1968b}, 
using the averaging procedure~\cite{Zala:1992,Zala:1993} (one can also use
the Isaacson's averaging procedure~\cite{Isaa:1968a,Isaa:1968b}, see also 
\cite{Zala:1996}) and with taking into account (\ref{polar}) brings 
the averaged Einstein equations in the form:
\begin{equation}
\label{av-ein-molec}
R^{(0)}_{\mu \nu} -\frac 12 g^{(0)}_{\mu \nu} R^{(0)} = 
- \kappa (T^{\rm (free)}_{\mu \nu} + 
T^{\rm (GW)}_{\mu \nu } + \frac{1}{2} Q_{\mu \rho \nu \sigma}{}^{; \rho \sigma}) ,
\end{equation}
where $T^{\rm (GW)}_{\mu \nu }$ is Isaacson's energy-momentum tensor of 
gravitational
waves~\cite{Isaa:1968a,Isaa:1968b}. All members in equation (\ref{av-ein-molec}) 
can be 
shown to be of the same order of magnitude ${\cal O} (1/L^2)$. The macroscopic 
equations
(\ref{av-ein-molec}) give the equations of motion for molecules\footnote{
Under assumption that the background metric in the left-hand side of 
(\ref{av-ein-molec}) 
represents a weak gravitational field on the flat background one can use the 
covariant derivatives with respect the metric in all relations instead of 
partial derivatives with respect to the flat metric.}
\begin{equation}
\label{eq-motion-molec}
T^{{\rm (free)} \mu \nu}{}_{; \nu} = 0 ,
\end{equation}
conservation of the energy-momentum of gravitational waves
\begin{equation}
\label{conserv-gw}
T^{{\rm (GW)}\mu \nu}{}_{; \nu} = 0 ,
\end{equation}
and an identity for the gravitational polarization
\begin{equation}
\label{ident-polar}
Q_{\mu \nu \rho \sigma}{}^{; \nu \sigma \mu} = 0 .
\end{equation}

The system of equations (\ref{av-ein-molec})-(\ref{ident-polar}) is 
underdetermined - there are 20 unknown components of the tensor of 
gravitational polarization. It is possible
to formulate two natural material relations. The first relation is between 
the traceless part of the quadrupole polarization tensor 
\begin{equation}
\label{polar-decomp}
\widetilde{Q}_{\mu \rho \nu \sigma} = {Q}_{\mu \rho \nu \sigma} - 
\frac{1}{4} g_{\mu \nu} P_{\rho \sigma} ,
\end{equation}
where $P_{\rho \sigma} = {Q}^{\mu}{}_{\rho \mu \sigma}$, and traceless 
energy-momentum tensor of gravitational waves 
\begin{equation}
\label{material-1}
\frac{1}{2} \widetilde{Q}_{\mu \rho \nu \sigma}{}^{; \rho \sigma} = 
\lambda T^{\rm (GW)}_{\mu \nu } ,
\end{equation}
where $\lambda=\lambda (x)$. Relation (\ref{material-1}) can be shown to be always 
valid 
in the geometrical optics limit. 

The second material relation connects the remaining part 
of the polarization tensor ${Q}_{\mu \rho \mu \sigma}$, its trace 
$P_{\rho \sigma}$, with
a projection of the curvature tensor on the world line of an observer 
(electric part of 
the curvature tensor)
\begin{equation}
\label{material-2}
P_{\rho \sigma} = \epsilon R^{(0)}_{\mu \rho \nu \sigma} u^\mu u^\nu ,
\end{equation}
where $u^\mu$ is the observer 4-velocity (4-velocity of the 
molecule centre of mass) and $\epsilon=\epsilon(x)$. The relation 
(\ref{material-2}) can be shown to lead to the correct expression for 
the 3-tensor of the average quadrupole gravitational moment (\ref{quadruple}) 
so that
\begin{equation}
\label{q-trace}
P_{\mu \nu} = (P_{00} = 0, P_{0i} = 0, P_{ij} = \langle G_{ij} \rangle N) .
\end{equation}
Then the material relation (\ref{gr-material}) can be recovered in the form
\begin{equation}
\label{gr-material-improv}
Q_{i 0 j 0} = \langle G_{ij} \rangle N = \epsilon_g R_{i 0 j 0}
\end{equation}
that gives $\epsilon = \epsilon_g$ with the gravitational dielectric constant
$\epsilon_g$ defined by (\ref{grav-dielectric-const}).

Thus the system of equations  (\ref{av-ein-molec})-(\ref{conserv-gw}), 
(\ref{material-1}), 
(\ref{material-2}) is fully determined and can be used to find the gravitational 
and
polarization fields for the macroscopic gravitating systems.

\acknowledgments

Roustam Zalaletdinov would like to thank Remo Ruffini for hospitality in ICRA where 
the work has been done in part.


\begin{thebibliography}{0}

\bibitem{Tava-Zala:1998}  \BY{Tavakol~R. \atque Zalaletdinov~R.}
\IN{Found. Phys.}{28}{1998}{307}. 

\bibitem{Shir-Fish:1962}  \BY{Shirokov~M.F. \atque Fisher~I.Z.}
\IN{Astron. Zh.}{39}{1962}{899} (in Russian) [English translation: \IN{Sov. Astron. -
A.J.}{6}{1963}{699}].

\bibitem{Scia:1971}  \BY{Sciama~D.W.}
\TITLE{Modern Cosmology} (CUP, Cambridge) 1971, Chapter 8. 

\bibitem{Elli:1984}  \BY{Ellis~G.F.R. } in \TITLE{General Relativity and
Gravitation}, edited by \NAME{Bertotti~B, de Felici~F. \atque Pascolini~A.}
(Reidel, Dordrecht) 1984. 

\bibitem{Zala:1992}  \BY{Zalaletdinov~R.M.} \IN{Gen. Rel. Grav.}{24}{1992}{1015}.

\bibitem{Zoto-Stoe:1992}  \BY{Zotov~N.V. \atque Stoeger~W.R} \IN{Class. Quantum Grav.} 
{9}{1992}{1023}.

\bibitem{Zala:1997}  \BY{Zalaletdinov~R.M.} \IN{Bull. Astron. Soc. India}{25}{1997}{401}.

\bibitem{Zala:1998} \BY{Zalaletdinov~R.M.} \IN{Hadronic J.}{21}{1998}{170}.

\bibitem{Yodz:1971}  \BY{Yodzis~P.} \IN{Inter. J. Theor. Phys.}{3}{1971}{331}.

\bibitem{Igna:1978}  \BY{Ignatiev~Yu.G.} in \TITLE{Gravitatsya i Teoriya 
Otnositel'nosti 
(Gravitation and Relativity Theory)}, Issue 14-15 (Kazan State University Press, Kazan)
1978, (in Russian).

\bibitem{Zala:1993}  \BY{Zalaletdinov~R.M.} \IN{Gen. Rel. Grav.}{25}{1993}{673}.

\bibitem{Zala:1996}  \BY{Zalaletdinov~R.M.} \IN{Gen. Rel. Grav.}{28}{1996}{953}.

\bibitem{Kras:1997}  \BY{Krasi\'nski~A.} \TITLE{Inhomogeneous Cosmological Models} 
(Cambridge University Press, Cambridge) 1997. 

\bibitem{Mars-Zala:1997}  \BY{Mars~M. \atque Zalaletdinov~R.M.} \IN{J. Math. Phys.}{38}
{1997}{4741}.

\bibitem{Zala:1996b} \BY{Zalaletdinov~R.M.} in \TITLE{Proceedings of the 7th
Marcel Grossmann Meeting on General Relativity}, Stanford, USA, July 1994, 
Part A, edited by \NAME{Jantzen~R.T. \atque Mac Keiser~G.} (World Scientific, 
Singapore) 1996, p. 394.

\bibitem{Bild-Futa:1991}  \BY{Bildhauser~S. \atque Futamase~T.} \IN{Gen. Rel. Grav.} 
{23}{1991}{1251}.

\bibitem{Lore:1916}  \BY{Lorentz~H.A.} \TITLE{The Theory of Electrons}, (Teubner,
Leipzig) 1916.

\bibitem{Pano-Phil:1962}  \BY{Panovsky~W.K.H. \atque Phillips~M.} \TITLE{Classical
Electricity and Magnetism} (Addison-Wesley, Reading) 1962.

\bibitem{deGr-Sutt:1972}  \BY{de Groot S.T. \atque Suttorp L.G.} \TITLE{Foundations
of Electrodynamics } (North-Holland, Amsterdam) 1972.

\bibitem{Jack:1975} \BY{Jackson~J.D.} \TITLE{Classical Electrodynamics} (John
Wiley \& Sons, New York) 1975.

\bibitem{Cher:1962}  \BY{Chernikov N.A} \IN{Dokl. Akad. Nauk SSSR}{144}{1962}{544}
[\IN{Soviet Physics-Doklady}{7}{1962}{428}].

\bibitem{Isra:1972}  \BY{ Israel~W.} in \TITLE{General Relativity},
edited by \NAME{O'Raifeartaigh~L.} (Clarendon Press, Oxford) 1972.

\bibitem{deGr:1969}  \BY{de Groot S.R.} \TITLE{The Maxwell Equations}
(North-Holland, Amsterdam) 1969.

\bibitem{Russ:1970} \BY{Russakoff~G.} \IN{Amer. J. Phys.}{38}{1188}{1970}.

\bibitem{Robi:1973}  \BY{Robinson F.N.H.} \TITLE{Macroscopic Electrodynamics}
(Pergamon Press, Oxford) 1973.

\bibitem{Bohr-Rose:1933}  \BY{Bohr~N. \atque Rosenfeld~L.} 
\IN{Mat.-fys. Medd. Dan. Vid. Selsk.}{12}{1933}{} [English translation in 
Selected Papers of L\'eon Rosenfeld, edited by \NAME{Cohen~R.S. \atque Stachel~J.J}
(D.Reidel, Dordrecht) 1979, p.~357].

\bibitem{DeWi:1962}  \BY{DeWitt B.S} in \TITLE{Gravitation: An introduction to
current research}, edited by \NAME{Witten~L.} (Wiley, New York, 1962), p. 266.


\bibitem{Szek:1971}  \BY{Szekeres~P.} \IN{Ann. Phys. (NY)}{64}{1971}{599}.

\bibitem{Pira:1962}  \BY{Pirani~F.A.E.} in \TITLE{Gravitation: Introduction to 
Current Research}, edited by \NAME{Witten L.}(Wiley, New York) 1962.

\bibitem{Szek:1966}  \BY{Szekeres~P.} \IN{J. Math. Phys}{7}{1966}{751}.

\bibitem{Kauf:1962}  \BY{Kaufman~A.N.} \IN{Ann. Phys. (NY)}{18}{1962}{264}.

\bibitem{Bel:1961}  \BY{Bel~L.} \IN{Ann. Inst. Henri Poincar\'e}{17}{1961}{37}.

\bibitem{Isaa:1968a}  \BY{Isaacson~R.A.} \IN{Phys. Rev.}{166}{1968}{1263}.

\bibitem{Isaa:1968b}  \BY{Isaacson~R.A.} \IN{Phys. Rev.}{166}{1968}{1272}.


\end{thebibliography}
\end{document}